\documentclass[conference]{IEEEtran}
\usepackage[utf8]{inputenc}
\IEEEoverridecommandlockouts
\usepackage{amsmath,amssymb,amsfonts}
\usepackage{algorithmic}
\usepackage{graphicx}
\usepackage{textcomp}
\usepackage{xcolor}
\usepackage{soul}
\usepackage{verbatim}
\usepackage{flushend} 
\usepackage{url}
\usepackage{hyperref}
\usepackage{multirow}
 \usepackage{booktabs}

\usepackage[pscoord]{eso-pic} 
\newcommand{\placetextbox}[3]{
    \setbox0=\hbox{#3}
    \AddToShipoutPictureFG*{
        \put(\LenToUnit{#1\paperwidth},\LenToUnit{#2\paperheight}){
            \vtop{{\null}\makebox[0pt][c]{#3}}}
        }
}
\placetextbox{.2}{0.055}{978-3-903176-67-6 © 2025 IFIP}

\begin{document}

\title{
Reinforcement Learning with Graph Attention for Routing and Wavelength Assignment with Lightpath Reuse \\

\thanks{This work was supported by the Engineering and Physical Sciences Research Council (EPSRC) grant EP/S022139/1 - the Centre for Doctoral Training in Connected Electronic and Photonic Systems - and EPSRC Programme Grant TRANSNET EP/R035342/1.}
}

\author{\IEEEauthorblockN{Michael Doherty}
\IEEEauthorblockA{Optical Networks Group\\
Electronic \& Electrical Eng. Dept.\\
\textit{University College London}\\
London, UK \\
michael.doherty.21@ucl.ac.uk}
\and
\IEEEauthorblockN{Alejandra Beghelli}
\IEEEauthorblockA{Optical Networks Group\\
Electronic \& Electrical Eng. Dept.\\
\textit{University College London}\\
London, UK \\
alejandra.beghelli@ucl.ac.uk}
}

\maketitle

\begin{abstract}
Many works have investigated reinforcement learning (RL) for routing and spectrum assignment on flex-grid networks but only one work has examined RL for fixed-grid with flex-rate transponders, despite production systems using this paradigm. Flex-rate transponders allow existing lightpaths to accommodate new services, a task we term routing and wavelength assignment with lightpath reuse (RWA-LR). We re-examine this problem case and present a thorough benchmarking of heuristic algorithms for RWA-LR, which are shown to have 6\% increased throughput when candidate paths are ordered by number of hops, rather than total length. We train an RL agent for RWA-LR with graph attention networks for the policy and value functions to exploit the graph-structured data. We provide details of our methodology and open source all of our code for reproduction. We outperform the previous state-of-the-art RL approach by 2.5\% (17.4~Tbps mean additional throughput) and the best heuristic by 1.2\% (8.5~Tbps mean additional throughput). This marginal gain highlights the difficulty in learning effective RL policies on long horizon resource allocation tasks. 
\end{abstract}



\section{Introduction and Motivation}
\label{sec:intro}

Reinforcement learning (RL) has been considered a promising solution method for resource allocation tasks on optical networks due to its potential to learn novel strategies. RL solutions have been shown to compute allocations in comparable time to heuristic algorithms while approaching the solution quality of exact methods such as integer linear programming (ILP) \cite{di_cicco_deep_2022}. This combination of fast allocation with potentially superior performance to heuristics has made RL the subject of much research for traffic allocation on dynamic networks \cite{chen_deeprmsa_2019,shimoda_mask_2021,tang_heuristic_2022,xu_deep_2022,cheng_ptrnet-rsa_2024}. Elastic optical networks have been the main focus of this research due to their potential for efficient spectrum utilization if the problem of spectral fragmentation can be avoided \cite{gerstel_elastic_2012}. 

However, flex-rate networks \cite{lord_flexible_2022} with a fixed 50 or 100GHz grid offer efficient use of spectrum without fragmentation and it is reported that many current production systems use this paradigm \cite{nevin_techniques_2022}. Flex-rate transponders allow additional data services to be accommodated on existing lightpaths if they have spare capacity, a task termed routing and wavelength assignment with lightpath reuse (RWA-LR). Additionally, incremental loading of the network with non-expiring services is considered a more realistic traffic model than dynamic traffic \cite{nevin_techniques_2022}. The unknown nature of future requests in the incremental loading setting, with the constraint that active services cannot be disrupted, make ILP impractical, therefore RL is of interest. 

So far only one paper, Nevin et al. from 2022 \cite{nevin_techniques_2022}, has investigated RWA-LR and used a simple feed-forward neural network with only 10M training samples. We therefore aim to investigate how much we can improve on this work by employing a sophisticated graph attention network (GAT) architecture \cite{velickovic_graph_2018}, to take advantage of the graph-structure of the problem, and 200M training samples for the final training run. This new scale of training is enabled by our GPU-based simulation framework, XLRON \cite{doherty_xlron_2024}, which we make openly available \cite{doherty_xlron_2023}. 

To establish the strongest benchmark to compare our RL results against, we evaluate K-Shortest Path First-Fit (KSP-FF) and First-Fit K-Shortest Paths (FF-KSP) heuristics and, crucially, compare their results when the candidate paths are ordered according to different sort criteria: 1) number of hops or 2) total length in km. Our trained RL agent improves on the strongest benchmark (KSP-FF with hops-ordering) and Nevin et al.'s RL solution.

Our work makes three novel contributions:

\begin{itemize}
    \item[1.] Thorough benchmarking of KSP-FF and FF-KSP with optimized path ordering, which shows hops-ordering to improve performance, with up to 6\% mean throughput increase in our case of study.
    \item[2.] Details of a new methodology for GAT-based RL applied to the RWA-LR problem with incremental traffic, giving new state-of-the-art results for network throughput, with 8.5 Tbps additional capacity compared to best benchmark.
    \item[3.] Release of all training and evaluation code from GPU-based framework, which enables RL training at a much larger scale than CPU-bound approaches.
\end{itemize}

\section{Previous Work}
Many works have examined RL for resource allocation in optical networks, but only a few have employed graph-based neural network architectures. Almasan et al. \cite{almasan_deep_2022} were the first to explore a message passing neural network (MPNN) with RL for routing in an optical network, which demonstrated improved performance compared to a theoretical fluid model and some capability to generalize to new topologies (with degraded performance). Xu et al. \cite{xu_deep_2022} used a graph convolutional neural network (GCN) as part of the policy and value functions in their actor-critic architecture for dynamic routing, modulation, and spectrum allocation (RMSA), and demonstrated 28\% improvement in blocking probability over KSP-FF and 22\% better than their implementation of DeepRMSA \cite{chen_deeprmsa_2019}. The GAT architecture was used in a recent work by Xiong et al. \cite{xiong_graph_2024} which demonstrated service blocking probability reduced from 2\% to 1\% compared to KSP-FF with K=5 on NSFNET for dynamic RMSA. The trend toward more advanced graph-based architectures is continued by Cheng et al. with their work PtrNet-RSA \cite{cheng_ptrnet-rsa_2024}, which employs a pointer network \cite{vinyals_pointer_2015} to select the nodes of the path and overcome the limitation of considering only k-shortest paths. Most of these works consider very similar problem settings (topologies and traffic models) and use the same benchmarks without investigating if they are the best benchmarks available. In this work, we establish the strongest heuristic benchmarks available (by testing different sort criteria for the candidate paths) to ensure our RL results are not trivial.

As discussed, only one previous work has studied the RWA-LR problem. Nevin et al. \cite{nevin_techniques_2022} made major contributions to the field by introducing RAW-LR and providing a detailed description of the techniques necessary to improve on their benchmarks, including invalid action masking and shortened training episodes. They demonstrate 18.4 Tbps increase in network throughput compared to their best benchmark (FF-KSP with K=5 and paths ordered by length). In this work we build on their recommendations and incorporate two improvements: we include the GAT architecture in our learning algorithm to exploit graph structure and we use our GPU-based simulation and training framework to enable training in a high-data regime.


\section{Network Model}
\label{sec:method}

We recreate the problem setting from Nevin et al. exactly, including the physical layer model. The paper uses a regular incoherent nonlinear interference Gaussian noise (GN) model for the physical layer, assuming transmission at Shannon rate and rectangular spectrum pulses with channel bandwidth equal to symbol rate. The point-to-point throughput is calculated at optimum launch power, with nonlinear interference treated as white Gaussian noise, and is valid for total modulated bandwidth up to combined C+L band \cite{shevchenko_maximizing_2022}. The Shannon capacity is calculated according to equation \ref{eq:c_path}, assuming Nyquist rate transmission. $R_s$ is the symbol rate, $NSR_i$ is the noise-to-signal ratio of the $i^{th}$ link of the path.

\begin{equation}
C_{path} = 2R_S \cdot \log_2\left(1 + \frac{1}{\sum_i NSR_i}\right)
\label{eq:c_path}
\end{equation}

For brevity, we do not include full details of the formulae for $NSR$ here and instead direct readers to Nevin et al. \cite{nevin_techniques_2022}. This model assumes transponders are capable of maximum spectral efficiency (SE), which flex-rate transponders can approximately achieve through sophisticated modulation techniques such as probabilistic constellation shaping. It also assumes fully-loaded links, a common assumption that ensures the SNR of allocated services is not impacted by future allocations. Overall, this physical layer model presents a realistic and conservative setting that accounts for multiple physical layer effects and is valid for 10~THz of bandwidth.

To implement the physical layer model in simulation, we calculate the path capacity for every candidate path between each node pair. When a lightpath is established on that path, it is designated the pre-calculated capacity, which is incrementally reduced with each service that is added to the lightpath. This corresponds to the transponder increasing the the spectral efficiency of its modulation format to increase the data rate. Each lightpath can therefore accommodate multiple services until the maximum capacity is reached.

We align with Nevin et al. and consider the NSFNET topology with bi-directional links\footnotemark, 10~THz total modulated bandwidth, 100 WDM channels with 100~GHz channel width, and non-expiring service requests of 100~Gbps that arrive sequentially (we term this "incremental loading"). We define an episode as 10,000 service requests and consider a uniform traffic matrix (each node pair has equal chance of being selected for a request).

\footnotetext{We note real-world networks are likely to have dual fiber pairs with dedicated directions of propagation, but we consider bi-directional links to align with previous work.}

\begin{figure*}[t]
    \centering
    \includegraphics[width=\linewidth]{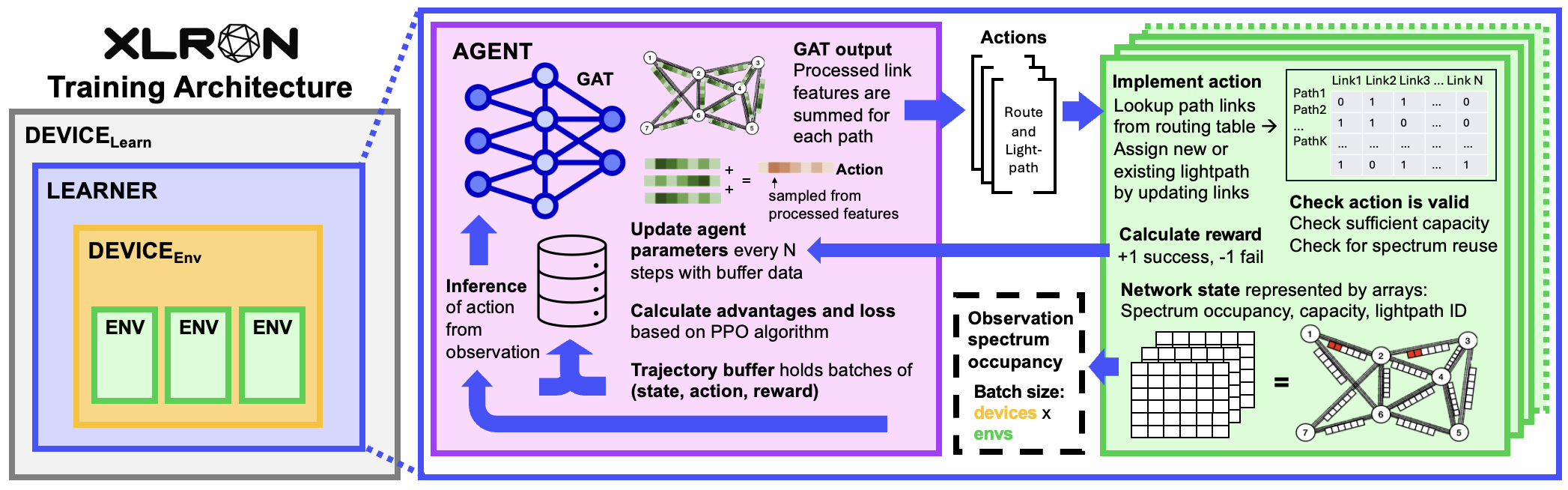}
    \caption{Outline of our RL training and optical network simulation framework, XLRON. The hierarchy of Device$_{LEARN}$, Learner, Device$_{ENV}$ and Environment computational abstractions are shown left. Right shows details of the training loop for a single Learner (set of neural network parameters) acting over parallel environments. The topology shown is illustrative; we use the NSFNET topology for our studies.}
    \label{fig:xlron}
\end{figure*}

We use the XLRON framework \cite{doherty_xlron_2023} to implement the RWA-LR environment. XLRON uses the JAX array computing framework \cite{bradbury_jax_2018} to allow compilation of programs to accelerator hardware, such as GPU or TPU. JAX imposes the constraint of functional programming to enable this compilation and requires all data structures used in XLRON are arrays. We track the spectrum occupancy, capacity of each lightpath, and the ID of each lightpath on every link. We verify that our recreation of Nevin et al.'s network model is accurate by comparison of our results for KSP-FF and FF-KSP, which reproduce Nevin et al.'s published results.

\section{Reinforcement Learning Agent Architecture}
\label{sec:rl}

We use the widely adopted Proximal Policy Optimization (PPO) \cite{schulman_proximal_2017} as our RL algorithm. PPO uses an actor-critic architecture, with separate learned policy and value functions. The value function provides a baseline estimate of the expected returns from a given state, which is subtracted from the calculated returns from a given rollout (sequence of state-action-reward tuples) to reduce the variance in the policy gradient, and give more stable updates to the parameters of the policy and value functions.

In our implementation, we use GAT's for the policy and value functions. GAT's use the message-passing framework of conventional graph neural networks with an additional learned attention mechanism to weight the contributions of messages from neighbors \cite{velickovic_graph_2018}. GAT's take graph-structured data as input and output. The input graphs for our GATs have the current service request as a global feature and the spectrum occupancy of links as edge features. The input graph features are processed by repeating the following steps for each round of message-passing: 1) Concatenate edge features with neighboring node features and global features, 2) transform concatenated features with a multi-layer perceptron (MLP), 3) apply a learned attention matrix to weight the transformed edge features, 4) aggregate the weighted edge features at nodes, 5) concatenate the node, global, and aggregated edge features, 6) transform the concatenated features with a MLP. 

The output graph features form the basis of our action selection distribution. For each candidate path, we element-wise sum the processed features of the path links, normalize the resulting array by the path length, and concatenate the aggregated arrays from each path to form a $K \times S$ array, where $K$ is the number of candidate paths and $S$ is the number of WDM channels. The array is passed to a softmax function to select the action, which is interpreted as a path and WDM channel. This method has the advantage of being scalable to an arbitrary number of paths. We also use the invalid action masking technique \cite{shimoda_mask_2021}. We make all code available on Github \cite{doherty_xlron_2023}.

Figure \ref{fig:xlron} provides an overview of the XLRON framework we use to implement our agent and environment. To enable scaling to hundreds of parallel environments on a single device, XLRON maintains a logical hierarchy of Device, Learner, and Environment, shown on the left of Figure \ref{fig:xlron}. A Learner represents a single learning agent, which can exist on one or more Devices. Environments can exist on only one Device and belong to one Learner, but the Learner may be replicated across Devices, to enable multi-device training. The right side of Figure \ref{fig:xlron} shows the RL training loop for a single Learner. The purple region indicates the internals of the agent with GAT for the policy and value functions, and rollout data stored in trajectory buffers for parameter updates. The batched parallel environments are illustrated in green, with each learner providing actions for the environments in parallel. The reward function for a single step is simply +1/-1 for success/failure in the service allocation.

\section{Training}
\label{sec:results}

The advantage of implementing both the agent and environment in JAX is the entire training loop can be compiled as a single program for greater efficiency, which is what enables us to scale training our computationally expensive GAT architecture to 200M samples. We perform all our experiments on a Nvidia A100 80GB. The training run took 10h 50m for 200M environment transitions across 100 parallel environments and a rollout length of 150, which equates to 13,333 parameter updates. Each parallel environment completed 1000 episodes of 2000 steps, giving 200M steps total. Training required 10GB of RAM. We could have scaled to more parallel environments to provide further speed-up but were limited by other users of the device. 

As recommended by Nevin et al., we use a scaling factor of 0.2 to reduce the length of our training episodes compared to our evaluation episodes. We train on episodes of 2,000 service requests and evaluate on 10,000. Each parallel environment therefore steps through 100,000 training episodes. This shorter training episode length reduces the horizon over which the RL agent needs to optimize, and reduces the difficulty of credit assignment \cite{pignatelli_survey_2023}, but is demonstrated to generalize to longer episodes in this case.

We performed a sweep of learning rate, discount factor, the generalized advantage estimation $\lambda$ parameter, GNN parameters of message passing steps, latent space dimensions, and learning rate schedule. We summarize the final values of the parameters in the following command, which we used to initiate the final training run:

{\footnotesize
\begin{samepage}
\begin{verbatim}
train.py --env_type rwa_lightpath_reuse
--incremental_loading --k 5 --link_resources 100
--topology_name nsfnet_deeprmsa_undirected  
--max_requests 10000 --values_bw 100 
--TOTAL_TIMESTEPS 200000000 --UPDATE_EPOCHS 10 
--ROLLOUT_LENGTH 150 --ACTION_MASKING 
--NUM_ENVS 100 --LR_SCHEDULE warmup_cosine 
--WARMUP_END_FRACTION 0.1 --scale_factor 0.2
--WARMUP_STEPS_FRACTION 0.1 --USE_GNN
--WARMUP_PEAK_MULTIPLIER 2 --gnn_mlp_layers 2
--message_passing_steps 3  --GAE_LAMBDA 0.984 
--GAMMA 0.919 --LR 1.943e-05 --gnn_latent 128 
\end{verbatim}
\end{samepage}
}

\begin{figure}[t]
    \centering
    \includegraphics[width=\linewidth]{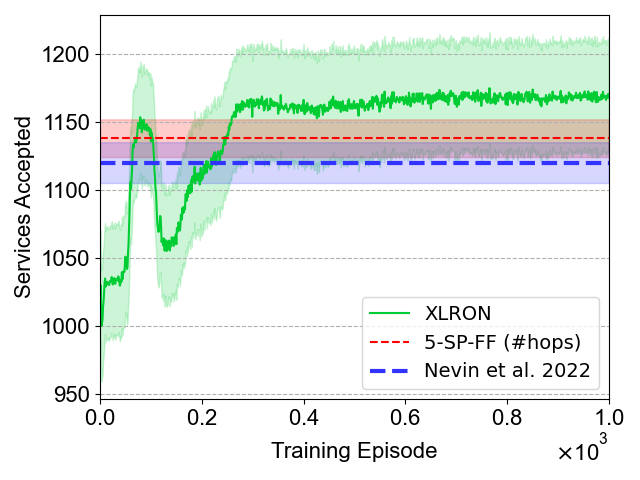}
    \caption{Training of our agent, compared with published RL results from Nevin et al. and our strongest heuristic benchmark (5-SP-FF with paths ordered by hops). Shaded areas indicate standard deviations. We used 100 parallel environments, with mean and standard deviation of accepted services at end of each episode calculated across environments.}
    \label{fig:training}
\end{figure}

Figure \ref{fig:training} shows the progression of training the XLRON agent in comparison to Nevin et al.'s published results for training, and the performance of the KSP-FF heuristic with paths ordered by hops. The agent exceeds the benchmark performance after training on 60M environment steps, after which it improves gradually and stabilizes.

\section{Evaluation}

\begin{figure}[t]
    \centering
    \includegraphics[width=\linewidth]{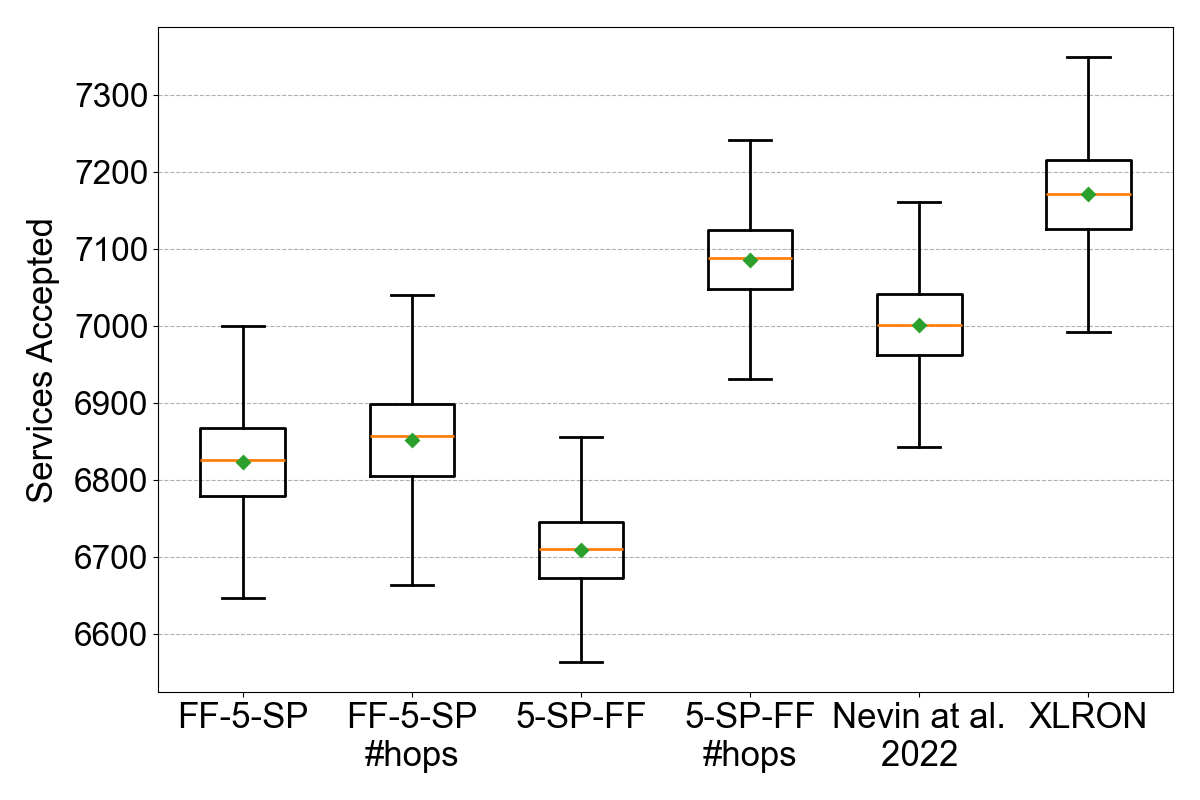}
    \caption{Each heuristic and RL solution was evaluated on 100 episodes fo 10,000 service requests. Boxplots show the mean, median, standard deviation and 1.5*interquartile range for accepted services from those episodes. \#hops indicates heuristics with candidate paths ordered by number of hops.}
    \label{fig:boxplots}
\end{figure}

To evaluate the trained model, we compare it to the best-performing heuristics from Nevin et al.: KSP-FF and FF-KSP \cite{vincent_scalable_2019}. We consider K=5 for our case to align with previous work. We include hops-ordering and length-ordering of paths in our analysis. Figure \ref{fig:boxplots} shows the distribution of total accepted services at the end of a 10,000 request evaluation episode, for 100 episodes in each case. Ordering by number of hops causes more requests to be accepted for both heuristics, with 6\% increased throughput for KSP-FF\footnotemark. This increase causes KSP-FF$_{hops}$ to exceed the RL results from Nevin et al., therefore highlighting the importance of benchmarking against diverse heuristics. XLRON improves on all cases, with 1.8\% increased throughput (8.5~Tbps) over KSP-FF$_{hops}$. We suggest this improvement is from to avoiding bottleneck links and better decisions to establish new lightpaths or use capacity on existing ones, which is made possible by learning from the high volume of simulated experience.

\footnotetext{We note that a more realistic physical layer model that accounts for ROADM losses at nodes is likely to show an even greater improvement in hops-ordering compared to length-ordering.}

\begin{figure}[t]
    \centering
    \includegraphics[width=\linewidth]{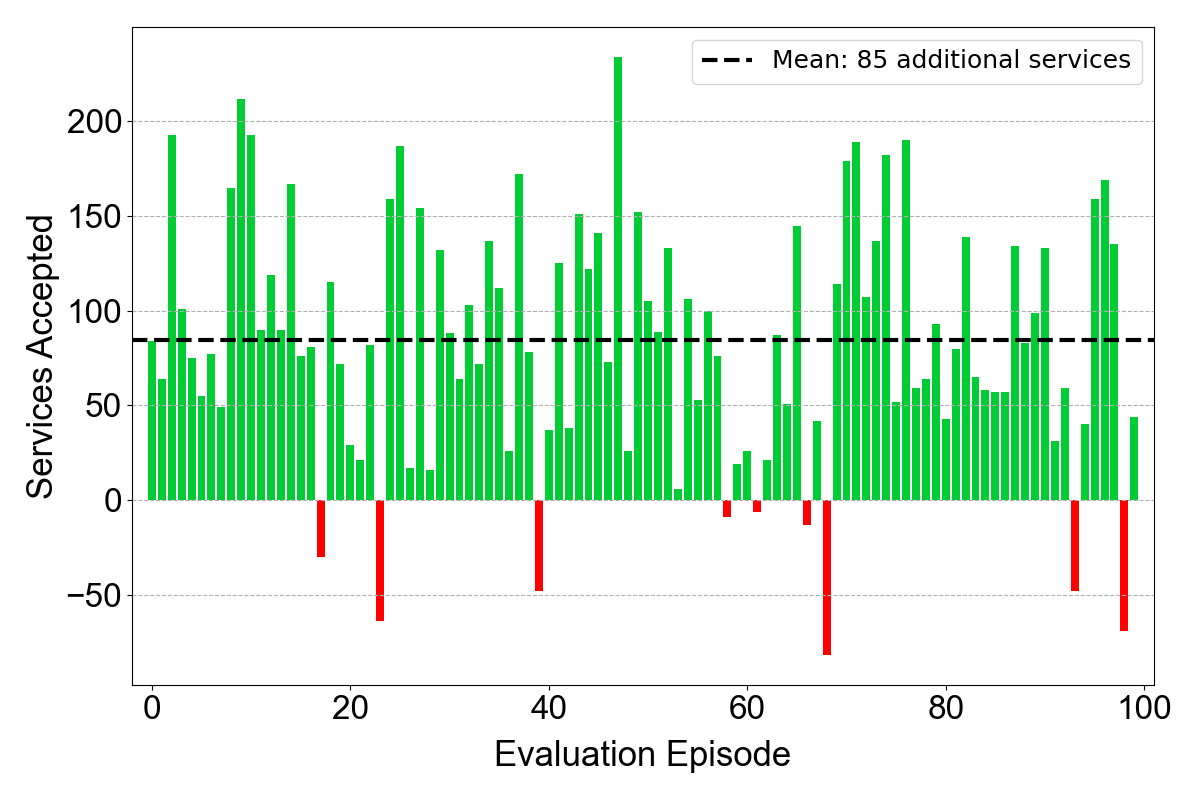}
    \caption{Comparison of services accepted by XLRON vs. KSP-FF$_{hops}$ for the same evaluation episodes (same sequences of service requests). Green bars indicate additional services accepted by XLRON, red indicates more services accepted by KSP-FF$_{hops}$.}
    \label{fig:waterfall}
\end{figure}

To thoroughly evaluate the performance of our agent, we apply both the trained agent and KSP-FF$_{hops}$ to the exact same sequence of traffic requests for 100 episodes. The difference in the total accepted services between the agent and heuristic in each episode is shown in Figure \ref{fig:waterfall}. In 91/100 episodes, our agent allows more services to be accepted, with a mean of 85 more. That the agent does not succeed in every episode shows the highly stochastic nature of the environment, for which it may be very difficult to learn a policy that outperforms in all cases.

We use K=5 in our experiments to enable direct comparison to  Nevin et al.'s results. However, we investigate if K=5 is the optimal number of candidate paths to consider, and how the episode length (number of service requests) affects the optimal K. We evaluate KSP-FF and FF-KSP for K=1 to K=10 for 100 evaluation episodes. We consider episode lengths of 10k, 15k, 20k, and 25k requests, and also consider the case of episode termination at the first blocking event. 

Table \ref{tab:comparison} shows the mean accepted services for each of these cases, with the best result at each episode length highlighted in bold. FF-KSP with K=2 accommodates the most services without a blocking event. KSP-FF with K=2 or 3 accepts more services for longer episode lengths. We conclude that incremental loading makes it favorable to consider a small number of candidate paths (instead of the commonly used K=5 for NSFNET), as this effectively rejects service requests that would otherwise be accommodated on longer more spectrally-expensive paths, which enables more less-expensive services to be accepted in total over the course of the episode. To avoid this, we suggest that future investigations of incremental loading seek to either maximize the target metric (e.g. throughput, energy use, revenue) until the first blocking event or minimize the resources required to accommodate all of a fixed number of requests.

\begin{table}[htbp]
\centering
\renewcommand{\arraystretch}{1.2}  
\begin{tabular}{c|c|ccccc}
\toprule
\multirow{2}{*}{Method} & \multirow{2}{*}{$K$} & \multicolumn{5}{c}{Episode Length} \\
\cmidrule{3-7}
& & 1st Blocking & 10k & 15k & 20k & 25k \\
\midrule
\multirow{10}{*}{\begin{tabular}{@{}c@{}} KSP-FF \\ \#hops \end{tabular}}
& 1 & 4309 & 6797 & 7830 & 8568 & 9191 \\
& 2 & 4435 & \textbf{7190} & 8146 & 8814 & 9349 \\
& 3 & 4805 & 7128 & \textbf{8172} & \textbf{9039} & \textbf{9774} \\
& 4 & 4864 & 7104 & 8134 & 9000 & 9725 \\
& 5 & 4876 & 7094 & 8119 & 8983 & 9685 \\
& 6 & 4902 & 7083 & 8107 & 8970 & 9668 \\
& 7 & 4916 & 7078 & 8101 & 8963 & 9661 \\
& 8 & 4931 & 7080 & 8098 & 8951 & 9646 \\
& 9 & 4940 & 7075 & 8093 & 8947 & 9634 \\
& 10 & 4972 & 7064 & 8080 & 8936 & 9614 \\
\midrule
\multirow{10}{*}{\begin{tabular}{@{}c@{}} FF-KSP \\ \#hops \end{tabular}}
& 1 & 4309 & 6797 & 7830 & 8568 & 9191 \\
& 2 & \textbf{5509} & 7091 & 8051 & 8647 & 8967 \\
& 3 & 5507 & 6953 & 7940 & 8653 & 9040 \\
& 4 & 5474 & 6899 & 7872 & 8581 & 8924 \\
& 5 & 5444 & 6849 & 7820 & 8477 & 8806 \\
& 6 & 5430 & 6827 & 7797 & 8445 & 8764 \\
& 7 & 5407 & 6810 & 7775 & 8385 & 8696 \\
& 8 & 5406 & 6789 & 7747 & 8340 & 8655 \\
& 9 & 5388 & 6778 & 7734 & 8313 & 8617 \\
& 10 & 5384 & 6762 & 7723 & 8295 & 8607 \\
\bottomrule
\end{tabular}
\vspace{0.2cm}
\caption{Comparison of mean services accepted by heuristics for different episode lengths and candidate paths (K).}
\label{tab:comparison}
\end{table}

\section{Conclusions}
\label{sec:conclusions}

Our work shows that it is possible to train an RL agent to outperform the best heuristic algorithms for RWA-LR. Thorough benchmarking is required to ensure that sophisticated RL solutions aren't outperformed by simple heuristics, and the ordering of shortest paths in heuristic algorithms has a large impact on throughput. The number of service requests in an episode also has a significant influence on the best heuristic and number of candidate paths.

Training an agent to outperform the best heuristic required considerable computational resources, engineering effort, and graph-based learning algorithms to take advantage of the graph structure of the data. We could not have achieved this without our GPU-based simulation framework, which enabled us to perform extensive hyperparameter tuning and a final training run 20X longer than the previous state of the art with the relatively compute-intensive GAT model architecture.

The less than 2\% increase in throughput we achieve highlights the difficulty of improving on the best heuristics and of learning effective RL policies on long horizon resource allocation tasks. We note that this gain is small in comparison to the projected annual increases in data traffic \cite{ericsson_ab_ericsson_2024}, and network operators might find more value in using RL to discover resource allocation policies that optimize other metrics, such as energy use \cite{martinez_enhancing_2024}.


\bibliography{paper.bib}

\end{document}